\documentclass[pra,reprint,a4paper,twocolumn,nofootinbib]{revtex4-2}

\usepackage{graphicx}
\usepackage{amsmath}
\usepackage{siunitx}
\usepackage{amssymb}
\usepackage{xcolor}
\usepackage[latin1]{inputenc}
\usepackage{booktabs}

\newcommand{\bichrom}[2]{($#1\omega$:$#2\omega$)}

\begin{document}

\title{Bichromatic phase-control of interfering Autler-Townes spectra}

\author{T. Bayer, K. Eickhoff, D. Köhnke and M. Wollenhaupt}
\affiliation{Carl von Ossietzky Universit\"at Oldenburg, Institut f\"ur Physik, Carl-von-Ossietzky-Stra\ss e 9-11, D-26129 Oldenburg Germany}

\date{\today}

\begin{abstract}
We propose a new scheme to control the shape of the Autler-Townes (AT) doublet in the photoelectron spectrum from atomic resonance-enhanced multiphoton ionization (REMPI). The scheme is based on the interference of two AT doublets created by ionization of the strongly driven atom from the ground and the resonantly excited state using tailored bichromatic femtosecond (fs) laser pulses. In this scheme, the quantum phase of the photoelectrons is crucial for the manipulation of the AT doublet. The laser polarization state and the relative optical phase between the two colors are used to manipulate the interference pattern. We develop an analytical model to describe the bichromatic REMPI process and provide a physical picture of the control mechanism. To validate the model, the results are compared to an \textit{ab initio} calculation based on the solution of the 2D time-dependent Schrödinger equation for the non-perturbative interaction of an atom with intense polarization-shaped bichromatic fs-laser pulses. Our results indicate that the control mechanism is robust with respect to the laser intensity facilitating its experimental observation. 
\end{abstract}

\maketitle

\section{Introduction} \label{sec:intro}

The use of strong laser fields to control ultrafast quantum dynamics enables efficient population transfer and opens up new excitation pathways due to the dynamic (AC) Stark effect. While the non-resonant AC Stark effect generally induces unidirectional energy shifts, the resonant AC Stark effect induces bidirectional shifts observed in the Autler-Townes (AT) splitting \cite{Autler:1955:PR:703}. Initially, phase-control of the AT doublet in the photoelectron energy spectrum from atomic (1+2) resonance-enhanced multiphoton ionization (REMPI) has been demonstrated using shaped single-color femtosecond (fs) laser pulses including pulse sequences \cite{Wollenhaupt:2003:PRA:015401,Wollenhaupt:2006:PRA:063409} and chirped pulses \cite{Wollenhaupt:2006:APB:183}. In these experiments, effective switching between the high-energy (fast) and low-energy (slow) AT components was observed. The control mechanism was shown to be the selective population of dressed states (SPODS) in the resonantly driven bound-system, mapped into the ionization continuum by the single-color driving field \cite{Bayer:2016:ACP:235}.\\
Here, we propose a new scheme to control the shape of AT spectra using tailored bichromatic fs-laser pulses with commensurable central frequencies. Adding a second color allows us to map not only the excited-state dynamics as in the previous single-color schemes, but also the ground-state dynamics \textit{via} direct multiphoton ionization (MPI). The frequency of the second color is chosen so that both AT doubles overlap in the photoelectron energy spectrum, resulting in an interference pattern in the superposition AT spectrum. In the following, the bichromatic control scheme is briefly referred to as IATS (interference of AT spectra). While the single-color SPODS schemes are based on locking the resonant bound-system in a state of maximum coherence \cite{Bai:1985:PRL:1277,Sleva:1985:JOSAB:483,Wollenhaupt:2003:PRA:015401}, the IATS scheme relies on Rabi oscillations between the resonantly coupled states. Due to the distinct relationship between the time-dependent amplitudes in the ground and excited states, described by the Rabi solution \cite{Rabi:1937:PR:652}, the photoelectron wave packets created \textit{via} the two different MPI pathways always interfere constructively in the slow AT component and destructively in the fast component. Very recently, the effect was observed in a single-color study of Rabi-dynamics in helium atoms using extreme ultraviolet femtosecond laser pulses from a free-electron laser \cite{Nandi:2022:Nature:488}. In the single-color scenario, the resonant and the non-resonant ionization pathway consist of the same number of photons. As a result, the created photoelectron partial wave packets have the same angular momentum state so that the effect can already be observed in the angle-integrated photoelectron energy spectrum \cite{Nandi:2022:Nature:488,Olofsson:2023:PRR:043017}. However, because both probe processes are driven by the same field, they are not controllable independently. In contrast, in the bichromatic IATS scheme introduced here, each process is driven by a different field component (color) and therefore  decoupled. Since each field component can be adjusted individually, the interference pattern is fully controlled by the relative optical phase between the two colors or their polarization state, which we will show below. Overlapping the AT spectra from both processes requires that the resonant and the non-resonant ionization pathway consist of a different number of photons.  As a consequence, the angular momentum states of the partial wave packets are generally different, which results in an angle-dependent interference. The slow AT component is selectively observed in certain directions, while the fast component is detected in other directions. In this case, the differential, i.e., energy- and angle-resolved measurement of the photoelectron momentum distribution is crucial to observe the effect. In contrast to the photon locking scheme, the interference pattern in the IATS scheme is independent of the field amplitude. This robustness of the interference mechanism to intensity fluctuations and focal intensity averaging will facilitate its observation in the experiment. \\
In this contribution, we present a combined analytical and numerical study of the novel IATS scheme. For the showcase example of a two-state atom perturbatively coupled to a photoionization continuum, we derive analytical expressions for the interference of AT spectra from REMPI with tailored bichromatic fs-laser pulses. The analytical model provides a clear physical picture of the interference mechanism and shows how it can be controlled by the optical phase and laser polarization. To validate the analytical model, the results are compared to \textit{ab initio} calculations based on the solution of the two-dimensional (2D) time-dependent Schrödinger equation (TDSE) for the non-perturbative interaction of a single-active-electron atom with polarization-tailored bichromatic fs-laser pulses \cite{Bayer:2020:PRA:013104,Bayer:2022:FC:899461}. The full calculation confirms the results of the analytical model and, in addition, reveals the influence of additional intermediate states on the multiphoton ionization (MPI) processes, which we recently discussed in more detail in \cite{Bayer:2023:PRA:033111}.

\section{Physical mechanism}\label{sec:mechanism}

\begin{figure*}[t]
	\includegraphics[width=1.0\linewidth]{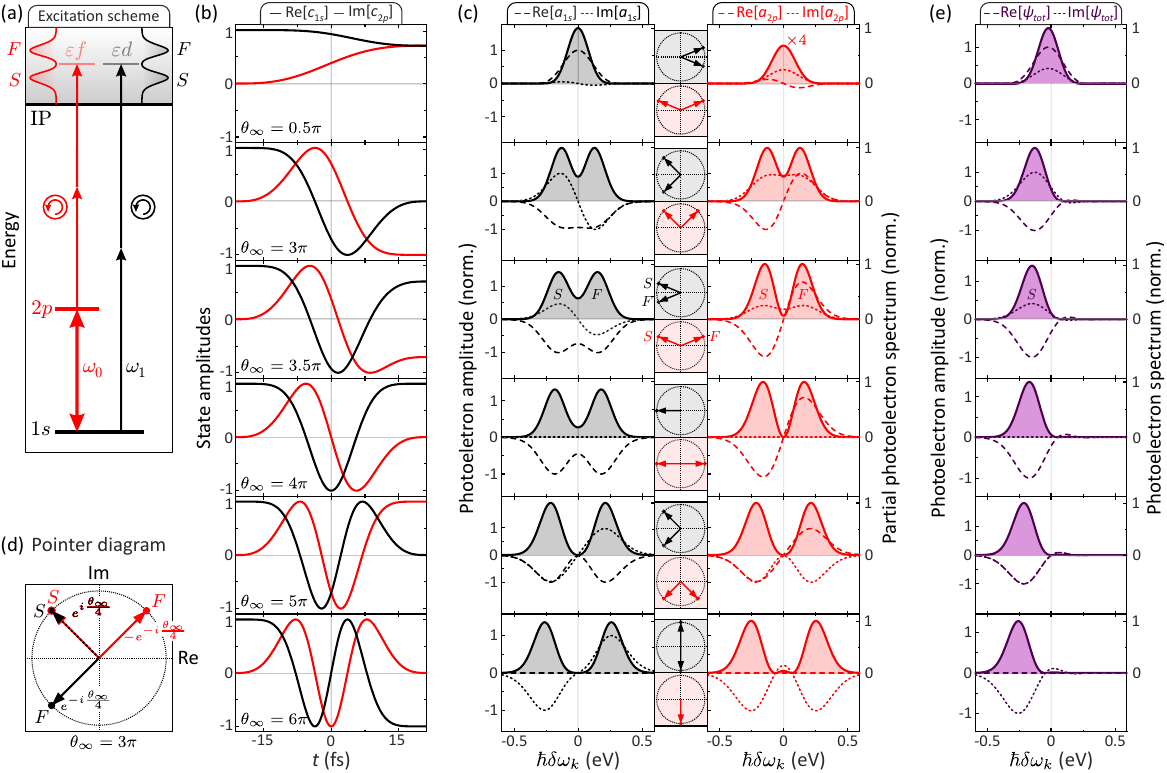}
	\caption{(Color online.) Illustration of the IATS scheme on the example of an (1+2) REMPI vs. non-resonant 2PI scenario. The corresponding excitation scheme is shown in (a). (b) shows the population amplitudes of the resonantly coupled bound-states $1s$ (black lines) and $2p$ (red lines) for different pulse areas $\theta_\infty$. The amplitudes of the created photoelectron partial wave packets are shown in (c) as dashed (real part) and dotted (imaginary part) lines, along with created partial AT spectra (solid lines). The phase relation between the respective slow (S) and fast (F) components of both spectra is indicated in the central column by the pointers and exemplified for $\theta_\infty=3\pi$ in (d). For all pulse areas, the slow components are in phase, whereas the fast components are in anti-phase. As a consequence, the fast components always interfere destructively and are eliminated from the total AT spectra displayed in (e).} 
	\label{fig:fig1}
\end{figure*}

\subsection{General scheme}\label{sec:mechanism:scheme}

We start by introducing the new scheme on a generic model system. To this end, we consider a two-state atom with states labeled by $1s$ and $2p$. The states have the eigenenergies $\varepsilon_{1s}=0$ and $\varepsilon_{2p}=\hbar\omega_{2p}$ and are dipole-coupled by the transition matrix element $\mu_0$. In addition, both states are coupled to an ionization continuum with ionization potential (IP) $\varepsilon_\text{IP}=\hbar\omega_\text{IP}$. Following the experimental scheme in \cite{Wollenhaupt:2003:PRA:015401}, we mimic the potassium atom by setting $\hbar\omega_{2p}=\SI{1.61}{eV}$ and $\hbar\omega_\text{IP}=\SI{4.34}{eV}$ \cite{Kramida:2018}. Initially, we investigate the interaction of the atom with a corotating circularly polarized (COCP) bichromatic laser field, which is described in the spherical basis by its positive frequency analytic signal as
\begin{equation}\label{eq:field}
	\boldsymbol{E}^+(t) = f(t)\big[\mathcal{E}_0e^{i\omega_0 t}+\mathcal{E}_1e^{i(\omega_1 t+\varphi)}\big]\boldsymbol{e}_{\pm1}.
\end{equation}
The two spectral field components are characterized by individual amplitudes $\mathcal{E}_j$ and central frequencies $\omega_j$ ($j=0,1$), a relative phase of $\varphi$ and a common temporal pulse envelope $f(t)$ with unit peak amplitude. The spherical unit vectors $\boldsymbol{e}_{\pm1}=(\boldsymbol{e}_x\mp i\boldsymbol{e}_y)/\sqrt{2}$ describe left- (LCP) and right-handed circularly polarized (RCP) light, respectively in the $x$-$y$-plane. The first field component is tuned to the atomic resonance by setting $\omega_0=\omega_{2p}$. By absorption of two additional photons $\hbar\omega_0$, the atom is ionized in a (1+2) REMPI process which maps the dynamics of the $2p$-state into an $f$-type photoelectron continuum. Simultaneously, the ground-state dynamics is mapped into a $d$-type continuum by non-resonant two-photon ionization (2PI) of the $1s$-state by the second field component. Setting $\omega_1=3\omega_0/2$, the created photoelectron wave packet overlaps energetically with that from the (1+2) REMPI process. The scheme is depicted in Fig.~\ref{fig:fig1}(a). We note, that according to our full TDSE calculations, additional photoelectron contributions due to non-resonant three-photon ionization from the $1s$ ground state by the $\omega_0$-component, analogous to those observed in \cite{Nandi:2022:Nature:488}, are negligible in our scheme and hence not considered in the analytical model. In the following, we discuss the analytical expressions for the interference of the two partial wave packets derived in the Appendix. \\
The photoelectron wave function can be expressed as the product of an energy-dependent amplitude $a_n(\omega_k)$ and an angular part $Y_{\ell,m}(\vartheta,\phi)$ \cite{Parfitt:2002:JMP:4681b}:
\begin{equation}\label{eq:photoe_wfct}
	\psi_{n,\ell,m}(\omega_k,\vartheta,\phi) = a_n(\omega_k)\,Y_{\ell,m}(\vartheta,\phi).
\end{equation}
Here, the index $n$ labels the bound state from which the ionization was initiated ($1s$ or $2p$) and $\varepsilon=\hbar\omega_k$ is the photoelectron kinetic excess energy. The indices $\ell$ and $m$ are the orbital angular momentum quantum numbers. In view of the comparison between the analytical model and our 2D-TDSE model, we restrict the description to the 2D case and drop the angular coordinate $\vartheta$, along with the corresponding quantum number $\ell$. The angular part then reduces to $Y_m(\phi)= e^{i m\phi}$, with $m=\pm2$ for the $d$-type wave packet from the $1s$ ground state and $m=\pm3$ for the $f$-type wave packet from the $2p$ excited state \cite{Parfitt:2002:JMP:4681a}. The plus and minus sign correspond to LCP and RCP ionization, respectively. Since the AT doublet manifests in the photoelectron energy spectrum, we initially discuss their amplitudes $a_n(\omega_k)$.
To this end, we apply second order time-dependent perturbation theory, using $\boldsymbol{E}^-(t)=[\boldsymbol{E}^+(t)]^\ast$, and make the ansatz \cite{Meier:1994:PRL:3207,Wollenhaupt:2003:PRA:015401}:
\begin{align}
 	a_{1s}(\delta\omega_k) & = \alpha_{1s,d} \intop_{-\infty}^{\infty} c_{1s}(t) \,  f^2(t) \, e^{i \delta\omega_k t} \,  dt \label{eq:ansatz_1s} \\
	a_{2p}(\delta\omega_k) & = \alpha_{2p,f} \intop_{-\infty}^{\infty} c_{2p}(t) \,  f^2(t) \, e^{i \delta\omega_k t} \,  dt, \label{eq:ansatz_2p}
\end{align}
where the $c_n(t)$ are the complex-valued bound-state amplitudes and 
\begin{equation}
	\delta\omega_k=\omega_k+\omega_\text{IP}-3\omega_0=\omega_k+\omega_\text{IP}-2\omega_1
\end{equation}
is the detuning of three photons $\hbar\omega_0$ -- or two photons $\hbar\omega_1$, due to the proper choice of $\omega_1$ -- from the continuum state $\hbar(\omega_k+\omega_\text{IP})$. The prefactors
\begin{equation}\label{eq:ion_amps}
	\alpha_{1s,d} = \gamma^{(2)}_{1s,d}\,\mathcal{E}_1^2\,e^{-i2\varphi}\quad \text{and} \quad \alpha_{2p,f} =\gamma^{(2)}_{2p,f}\,\mathcal{E}_0^2,
\end{equation}
with $\gamma_{n,m}^{(2)}$ describing the two-photon coupling of the bound-states to the respective continua, are matched by suitable choice of the field amplitudes $\mathcal{E}_j$. A constant relative phase of the $\gamma_{n,m}^{(2)}$ can be compensated by an additional relative optical phase. In the following, we therefore assume $\alpha_{1s,d}=\alpha_{2p,f}$. In the resonant case, the bound-state amplitudes are given by the Rabi solution \cite{Rabi:1937:PR:652,Shore:1990a} in Eq.~\eqref{eq:Rabi_app} of the Appendix. Using a cosine-squared pulse envelope with a footprint duration of $\Delta t$ and a pulse area of $\theta_\infty$  (see Eqns.~\eqref{eq:envelope} and \eqref{eq:pulsearea}), the integrals in Eqns.~\eqref{eq:ansatz_1s} and \eqref{eq:ansatz_2p} can be solved analytically. The derivation for the general case of $(1+N)$ REMPI vs. $M$-photon ionization is given in the Appendix (Sec.~\ref{app:analytics}). For the specific case of (1+2) REMPI vs. 2PI, the photoelectron amplitudes take the form:
\begin{align}
	a_{1s}(\delta\omega_k) & \propto  \mathcal{A}_2(\delta\omega_k) +  \mathcal{A}_2^*(-\delta\omega_k) \label{eq:solution_1s} \\
	a_{2p}(\delta\omega_k) & \propto   \mathcal{A}_2(\delta\omega_k) - \mathcal{A}_2^*(-\delta\omega_k) \label{eq:solution_2p}.
\end{align}
The function
\begin{equation}\label{eq:Anger}
	\mathcal{A}_2(\delta\omega_k) = \frac{\Delta t}{64} \, e^{i \frac{\theta_\infty}{4} } \, \sum_{j=0}^4 \begin{pmatrix}
		4 \\ 
		j
	\end{pmatrix} \, \boldsymbol{\mathrm{J}}_{\nu_j(\delta\omega_k)}\left(\frac{\theta_\infty}{4\pi} \right)
\end{equation}
describes a superposition of real-valued Anger functions $\boldsymbol{\mathrm{J}}_{\nu_j}$ of the order 
\begin{equation}
\nu_j (\delta\omega_k) = -\left(2-j + \frac{\theta_\infty}{4\pi}+\frac{\delta\omega_k\Delta t}{2\pi}\right).
\end{equation}
The order is a continuous parameter which determines the spectral position of the Anger functions (see Fig.~\ref{fig:app1} in the Appendix for an illustration). We note that Eqns.~\eqref{eq:solution_1s} and \eqref{eq:solution_2p} are not only valid for cosine-squared pulses but hold true for any pulse with a real-valued envelope irrespective of its shape.
\\
To illustrate the analytical results of Eqns.~\eqref{eq:solution_1s} and \eqref{eq:solution_2p}, we show photoelectron amplitudes for different pulse areas $\theta_\infty$, ranging from $0.5\pi$ to $6\pi$, in  Fig.~\ref{fig:fig1}. The pulse duration is set to $\Delta t=\SI{41.2}{fs}$ which corresponds to a full width at half maximum (FWHM) of the intensity $f^2(t)$ of $\Delta t_\mathrm{fwhm}=\SI{15}{fs}$, as used in the \textit{ab initio} calculation in Secs.~\ref{sec:results:COCP} and \ref{sec:results:CRCP}. The underlying bound-state dynamics are shown in Fig.~\ref{fig:fig1}(b). The ground-state amplitude $c_{1s}(t)$ (black line) and the excited-state amplitude $c_{2p}(t)$ (red line) display Rabi cycling, except for the weak-field scenario $\theta_\infty=0.5\pi$ shown in the first row. Fig.~\ref{fig:fig1}(c) shows the photoelectron amplitudes $a_{1s}(\delta\omega_k)$ calculated using Eq.~\eqref{eq:solution_1s} (left column, black lines) and $a_{2p}(\delta\omega_k)$ calculated using Eq.~\eqref{eq:solution_2p} (right column, red lines), decomposed into real (dashed) and imaginary part (dotted), along with the corresponding energy spectra $|a_n(\delta\omega_k)|^2$ (solid shaded). For pulse areas $\theta_\infty\gtrsim3\pi$, the latter reveal a distinct AT splitting in both partial wave packets. The slow AT components ($S$) are related to the first term in Eqns.~\eqref{eq:solution_1s} and \eqref{eq:solution_2p} while the fast components ($F$) are associated with the second term. The shape of each AT component is determined by the function $\mathcal{A}_2(\delta\omega_k)$ in Eq.~\eqref{eq:Anger}, which describes a superposition of real-valued Anger functions with a common phase of $\theta_\infty/4$ for the slow and $-\theta_\infty/4$ for the fast components. The amplitudes are therefore real-valued for pulse areas of even multiples of $2\pi$, e.g., for $\theta_\infty=4\pi$ as shown in the fourth row. For odd multiples of $2\pi$, they are purely imaginary, e.g., for $\theta_\infty=6\pi$ as shown in the bottom row. The key result, however, is the additional minus-sign in Eq.~\eqref{eq:solution_2p}, which arises due to the sinusoidal time-dependence of the excited-state amplitude $c_{2p}(t)$ in the Rabi solution, in contrast to the cosinusoidal behavior of $c_{1s}(t)$. Consequently, the fast components of both AT doublets are in anti-phase with each other, while the slow components are in-phase. To illustrate the phase relation of the AT components, the total phase of each component is indicated by pointers in the central column of Fig.~\ref{fig:fig1}(c). A more detailed pointer diagram is depicted in Fig.~\ref{fig:fig1}(d) for clarification, including an assignment of the phase factors $\pm e^{\pm i\frac{\theta_\infty}{4}}$ exemplarily for the pulse area $\theta_\infty=3\pi$. Considering the full photoelectron wave function in the direction $\phi=0$, where the angular parts of both partial wave packets are equal, the two wave packets \textit{always interfere constructively in the slow and destructively in the fast AT component}. As a result, the fast AT component is completely suppressed in the total photoelectron wave function in this direction:
\begin{align}
	\psi_{tot}(\omega_k,\phi=0) & = \psi_{1s,d}(\omega_k,0)+\psi_{2p,f}(\omega_k,0) \notag \\
	& = a_{1s}(\delta\omega_k) + a_{2p}(\delta\omega_k) \notag \\
	& \propto 2\mathcal{A}_2(\delta\omega_k).
\end{align}
Sections through the total photoelectron energy distribution (PED) $\mathcal{P}(\omega_k,\phi)=|\psi_{tot}(\omega_k,\phi)|^2$ along $\phi=0$ are displayed in Fig.~\ref{fig:fig1}(e) (purple solid line). The spectra clearly show the selective emission of slow AT electrons for all pulse areas $\theta_\infty$. This remarkable property will be advantageous for the experimental implementation of the IATS scheme, since it makes the scheme robust against focal intensity averaging which generally complicates the observation of strong-field phenomena.

\subsection{Optical control}\label{sec:mechanism:control}

In the (1+2) REMPI vs. 2PI scenario using \bichrom{2}{3} COCP pulses discussed so far, the interference of the two partial wave packets is angle-dependent, due to the different azimuthal phases $e^{\pm i3\phi}$ and $e^{\pm i2\phi}$ of their angular momentum states. In the opposite direction, i.e. for $\phi=\pi$, the wave function $\psi_{2p,f}$ acquires an additional phase of $\pi$ relative to the wave function $\psi_{1s,d}$. The additional sign inverts the interference condition and switches the photoelectrons selectively to the fast AT component:
\begin{align}
	\psi_{tot}(\omega_k,\pi) & = a_{1s}(\delta\omega_k) - a_{2p}(\delta\omega_k) \notag \\
	& \propto 2\mathcal{A}^*_2(-\delta\omega_k).
\end{align}
Besides the number of absorbed photons, the angular momentum states are determined by the polarization state of the two field components. This provides a handle to control the interference between the two partial wave packets. For example, switching the circularity of the second field component, thus generating a counterrotating circularly polarized (CRCP) \bichrom{2}{3} field, the angular parts of the partial wave packets become $e^{\pm i3\phi}$ and $e^{\mp i2\phi}$, respectively. The angle-dependent relative phase between the two wave functions $\psi_{2p,f}$ and $\psi_{1s,d}$ then equals $e^{\pm i5\phi}$, which inverts the interference condition in the directions $\phi_j=(2j+1)\cdot\pi/5$, ($j=0,1,..,4$). \\
The canonical parameter to control the interference, however, is the relative phase $\varphi$ between the field components (see Eq.~\eqref{eq:field}). This optical phase enters as a phase of $2\varphi$ (due to the two-photon ionization) into the amplitude $a_{1s,d}(\delta\omega_k)$ through the prefactor $\alpha_{1s,d}$ in Eq.~\eqref{eq:ion_amps}. Accordingly, a relative phase of $\varphi=\pi/2$ inverts the interference for all angles $\phi$. This type of optical \textit{phase control} of the AT doublet is studied in Sec.~\ref{sec:results:COCP}. The optical \textit{polarization control} of the AT doublet is studied in Sec.~\ref{sec:results:CRCP}.

\section{Results}\label{sec:results}

The results are presented in three parts. In Sec.~\ref{sec:results:COCP}, we demonstrate the new AT control scheme using COCP pulses and exert phase control on the AT doublet by the relative optical phase. Then, we study optical polarization control of the AT doublet using CRCP pulses in Sec.~\ref{sec:results:CRCP}. Finally, in Sec.~\ref{sec:results:LP}, we discuss the influence of additional MPI pathways which arise when linearly polarized pulses are used. In all sections, we employ different levels of theory. The analytical model introduced in Sec.~\ref{sec:mechanism} is validated against \textit{ab initio} calculations presented in the Appendix~\ref{app:ab_initio}. Deviations between both approaches are analyzed using multistate model simulations building a bridge between the generic two-state model and the full TDSE calculation.

\subsection{Corotating circularly polarized pulses}\label{sec:results:COCP}

\begin{figure}[t!]
	\includegraphics[width=\linewidth]{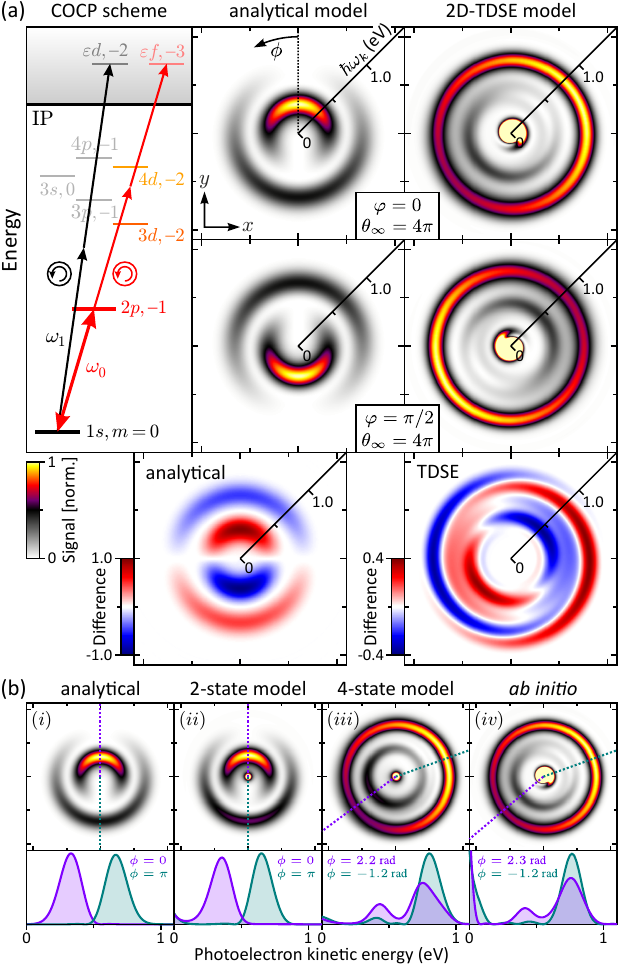}
	\caption{(Color online.) Implementation of the IATS scheme in a 2D potassium-like atom by COCP \bichrom{2}{3} pulses. (a) The left frame shows the excitation scheme for RCP pulses driving $\Delta m=-1$ transitions. The superposition of a $|d,-1\rangle$- and a $|f,-3\rangle$-type photoelectron partial wave packet gives rise to a $c_1$-symmetric 2D PED $\mathcal{P}(\omega_k,\phi)$. The PEDs from the analytical model are shown in the central column and compared to the numerical PEDs from the 2D-TDSE model in the right column. The top row shows results for an optical phase of $\varphi=0$. Below, the PEDs obtained for $\varphi=\pi/2$ are shown, demonstrating optical phase-control. The bottom row displays the difference between the two normalized PEDs from each model. (b) To analyze the influence of additional intermediate states in the \textit{ab initio} calculation, we employed different multi-state model simulations. While the two-state model in frame $(ii)$ reproduces the analytical results in $(i)$ accurately, reproduction of the full TDSE calculation in $(iv)$ requires inclusion of the intermediate states $3d$ and $4d$ (frame $(iii)$). The bottom row displays energy-resolved photoelectron spectra, similar to Fig.~\ref{fig:fig1}(e), taken in different directions $\phi$ indicated by dotted lines in the 2D PEDs.}
	\label{fig:fig2}
\end{figure}
In this section, we demonstrate the IATS scheme introduced in Sec.~\ref{sec:mechanism:scheme}. Using COCP \bichrom{2}{3} pulses with an FWHM duration of $\Delta t_\mathrm{fwhm} =\SI{15}{fs}$ and a pulse area of $\theta_\infty=4\pi$, we calculate the 2D PED $\mathcal{P}(\omega_k,\phi)$ according to Eq.~\eqref{eq:2D_PED} in Appendix~\ref{app:analytics}. The analytical PEDs are compared to those obtained from the 2D-TDSE model introduced in \cite{Bayer:2020:PRA:013104} (see also Appendix~\ref{app:ab_initio}). The results are presented in Fig.~\ref{fig:fig2}(a). The left frame shows the excitation scheme for the interaction of the atom with bichromatic RCP pulses driving $\Delta m=-1$ transitions. Due to the associated decrease of $m$ with every absorbed photon, the dipole selection rules allow only transitions with $\Delta \ell=+1$ in this case. Therefore, only the 2PI and the (1+2) REMPI pathway indicated in Fig.~\ref{fig:fig2}(a) are dipole-allowed. Each field component addresses only a single $\ell$-continuum, which is crucial for the background-free mapping of the bound-state dynamics as we will discuss in Sec.~\ref{sec:results:LP}. The analytical PED takes the form 
\begin{align}\label{eq:PED_COCP}
	\mathcal{P}(\omega_k,\phi)  & = 2\vert\mathcal{A}_2(\delta\omega_k)\vert^2 \bigg[1+\cos\left(\phi+2\varphi\right)\bigg] \notag \\ 
	& \quad +2\vert\mathcal{A}_2(-\delta\omega_k)\vert^2\bigg[1+\cos\left(\phi+\pi+2\varphi\right)\bigg].
\end{align}
A visualization of $\mathcal{P}(\omega_k,\phi)$ for $\varphi=0$ is shown in the central column (top frame) of Fig.~\ref{fig:fig2}(a). The PED displays two concentric crescents aligned in $y$-direction and rotated against each other by $\pi$. The inner crescent is described by the first term in Eq.~\eqref{eq:PED_COCP} and corresponds to the slow AT component in the photoelectron energy spectrum. The fast AT component corresponds to the outer crescent which is described by the second term in Eq.~\eqref{eq:PED_COCP}. The extra phase of $\pi$ in this second term reflects the minus sign in Eq.~\eqref{eq:solution_2p}. Along $\phi=0$, therefore only the slow AT component is observed while the fast component is completely suppressed. This is the scenario discussed in Sec.~\ref{sec:mechanism:scheme} and illustrated in Fig.~\ref{fig:fig1}. The corresponding energy-resolved spectrum along $\phi=0$ is shown in Fig.~\ref{fig:fig2}(b) as violet line in panel $(i)$. Along $\phi=\pi$, the AT doublet in the PED is inverted. Here the fast AT component is observed selectively while the slow component vanishes, in accordance with the discussion in Sec.~\ref{sec:mechanism:control}. The corresponding energy-resolved spectrum is shown as turquoise line in panel $(i)$ of Fig.~\ref{fig:fig2}(b). \\
Equation~\eqref{eq:PED_COCP} also shows that by introducing an optical phase of $\varphi$, the entire PED is rotated by an angle of $2\varphi$. Hence, setting $\varphi=\pi/2$ inverts the picture, as shown in the central frame of Fig.~\ref{fig:fig2}(a). Now the fast AT component is observed selectively along $\phi=0$ while the slow component is emitted selectively along $\phi=\pi$. \\
Next we compare the analytical 2D PEDs to numerical results from the 2D-TDSE model applied to the non-perturbative interaction of a 2D potassium-like atom with Gaussian-shaped COCP \bichrom{2}{3} pulses. A detailed characterization of the atomic system was provided in \cite{Bayer:2023:PRA:033111}. The field amplitudes of the two colors were set to $\mathcal{E}_0=\SI{2.7e7}{V/cm}$ and $\mathcal{E}_1=\SI{7.5e7}{V/cm}$, respectively, corresponding to peak intensities of $I_0=\SI{9.4e11}{W/cm^2}$ and $I_1=\SI{7.5e12}{W/cm^2}$. The full calculation results are displayed in the right column of Fig.~\ref{fig:fig2}(a). The numerical PEDs agree qualitatively with the analytical results. We observe the slow and fast AT components as two concentric rings with pronounced angular asymmetries. While the slow AT component is in fact crescent-shaped, as in the analytical model, the fast component appears enhanced by an isotropic offset which reduces its asymmetry contrast. In addition, both components are rotated by an angle of about $3\pi/4$ relative to the analytical results. A closer inspection reveals that the fast AT component exhibits a slight additional angular shift of about \SI{-0.36}{rad}. These rotations indicate the acquisition of different ionization phases which we will address in more detail below. Energy-resolved spectra along the directions where the slow and fast AT components are observed with maximum amplitude are shown as violet and turquoise line, respectively, in panel $(iv)$ of Fig.~\ref{fig:fig2}(b). The saturated contribution in the center of the PEDs results from frequency mixing between the two field components, i.e., the absorption of one photon $\hbar\omega_0$ and one photon $\hbar\omega_1$. The corresponding signal is centered around \SI{-0.30}{eV} below the IP but extends over the threshold due to the large spectral bandwidth of the colors (the FWHM of the frequency mixing contribution is \SI{0.26}{eV}). Most importantly, however, the maxima of the two AT components are aligned approximately in opposite directions, which is the signature of the IATS scheme. Also the inversion of the PED achieved by varying the optical phase to $\varphi=\pi/2$ is reproduced, as seen in the central frame of the right column in Fig.~\ref{fig:fig2}(a). To eliminate the isotropic offset observed in the fast AT component of the TDSE calculation, we calculated the difference between the two PEDs. The resulting differential PED is shown in the bottom row of Fig.~\ref{fig:fig2}(a) and compared to the result from the analytical model. The differential representation highlights the signatures of the IATS scheme even more clearly and reveals the significant degree of control, quantified by an asymmetry contrast of 40\%, in the full calculation. In addition, the differential PED from the 2D-TDSE model exhibits vortex structures, most pronounced in the fast AT component, which are indicative of linear spectral ionization phases induced by the intermediate $d$-resonances \cite{Eickhoff:2022:PRA:053113}.
\\
\begin{table}
	\setlength{\tabcolsep}{4pt}
	\begin{tabular}{c|cc|cc}
				&  \multicolumn{2}{c|}{COCP scenario} & \multicolumn{2}{c}{CRCP scenario}	\\
	transition	& \multicolumn{2}{c|}{$\mu_{nm}$ $(ea_0)$} & \multicolumn{2}{c}{$\mu_{nm}$ $(ea_0)$}  \\
		\hline
		$1s\rightarrow 2p$ & \multicolumn{2}{c|}{2.86}   & \multicolumn{2}{c}{2.86}   \\
		$2p\rightarrow 3d$ & \multicolumn{2}{c|}{3.31}   & \multicolumn{2}{c}{3.31}   \\
		$2p\rightarrow 4d$ & \multicolumn{2}{c|}{0.58}   & \multicolumn{2}{c}{0.58}   \\
		\hline
& amplitude	&  phase $[\pi]$	& amplitude	&  phase $[\pi]$	\\
		\hline
		$1s\rightarrow\rightarrow \varepsilon d$ & 0.30	& 0.81 & 0.30 & 0.60 \\
		$2p\rightarrow\rightarrow \varepsilon f$ & 1.00	& 0.59 & 0.58 & 1.22 \\
		$3d\rightarrow \varepsilon f$ & 0.03	 & 0.03 & 0.05 & 1.41 \\
		$4d\rightarrow \varepsilon f$ & 0.21	 & 1.48 & 0.17 & 0.43 \\
	\end{tabular}
	\caption{Upper part: Dipole matrix elements $\mu_{nm}=|\langle \psi_n|\boldsymbol{\mu}|\psi_m\rangle|$ for the most relevant bound-bound transitions of the 2D-TDSE model (cf. \cite{Bayer:2023:PRA:033111}). Lower part: Relative ionization amplitudes for the perturbative bound-ionic transitions considered in the four-state model.}
	\label{tab:couplings}
\end{table}
The deviations of the numerical results from the analytical model are attributed to the influence of additional intermediate states in the 2D-TDSE model. Specifically, the high-lying states $3d$, $3s$ and $4d$, indicated in the excitation scheme, were identified in \cite{Bayer:2023:PRA:033111} to play prominent roles in the non-perturbative interaction of the atom with intense LP laser pulses. In the COCP case, however, the $3s$-state is not accessible, due to the selection rule $\Delta\ell=+1$ mentioned above, and is therefore disregarded here. By analysis of the bound-state population dynamics, described in detail in \cite{Bayer:2023:PRA:033111}, we find that the blue-detuned intermediate state $4d$ is responsible for the enhancement of the fast AT component. Due to the resonant dynamic Stark effect in the strongly driven $1s$-$2p$ subsystem, the $4d$-state shifts into resonance with the $\omega_0$-field component. The detailed mechanism behind this enhancement was discussed in a dressed state picture in \cite{Bayer:2023:PRA:033111}. The rotation of the numerical PED relative to the analytical result is induced by ionization phases of two different types. The first type are static phases associated with different neutral-to-ionic couplings included in the $\gamma^{(2)}_{n,m}$ (see Eq.~\eqref{eq:ion_amps}) for the 2PI and the (1+2) REMPI pathway. In the 2D-TDSE calculation, such phases are inherently built-in. They can be compensated by a constant optical phase (provided the continuum is sufficiently flat in the relevant energy region). The second type are dynamic phases acquired in the REMPI pathway due to the transient resonance of the $4d$-state \cite{Eickhoff:2022:PRA:053113}. Such phases are responsible for the observed differential rotation of the fast AT component relative to the slow component. \\
To verify the analysis of the TDSE results and clarify the role of the intermediate states, we performed numerical multistate model simulations, similar to the strategy pursued in \cite{Bayer:2023:PRA:033111}. Starting from the two-state model, including only the ground state $1s$ and the resonant excited state $2p$, we successively included the additional states into the model and examined their influence on the bound-state dynamics and the PED. The dipole matrix elements for the bound-bound transitions from Ref.~\cite{Bayer:2023:PRA:033111} are provided in Tab.~\ref{tab:couplings}, along with the relative ionization amplitudes for the perturbative bound-ionic transitions used in the current four-state simulations. The results are shown in Fig.~\ref{fig:fig2}(b). The two-state model in panel $(ii)$ reproduces the PED from the analytical model in panel $(i)$ almost exactly. The only deviation is the frequency mixing contribution in the center of the numerical result, which is not captured in the analytical model. The overlap of the frequency mixing contribution with the slow AT component is sufficient to alter (reduce) the peak amplitude of the latter slightly. This is best discernible in the energy-resolved spectra displayed in the bottom frame of panel $(ii)$. Because both PEDs are normalized to their maximum, the fast AT component therefore appears slightly brighter in the numerical PED compared to the analytical PED. By incorporating the intermediate states $3d$ and $4d$ into the multistate model and adapting their complex-valued one-photon ionization (1PI) amplitudes (see Tab.~\ref{tab:couplings}), the simulation result in panel $(iii)$ is brought into very good agreement with the full calculation, shown again in panel $(iv)$. In particular, the PED from the refined four-state model displays an enhanced fast AT component, which we trace back to 1PI from the excited $4d$-state by the $\omega_0$-field. The rotation of the PED is due to the relative phase between two-photon ionization amplitudes $\gamma^{(2)}_{1s,-2}$ and $\gamma^{(2)}_{2p,-3}$. We even observe a small additional rotation of the fast AT component relative to the slow component, which is also induced by 1PI from the $4d$-state. In contrast, the influence of the red-detuned intermediate state $3d$ on the PED is only subtle. \\
The differences between the analytical and the numerical PEDs are also related to the population dynamics in the resonant bound-system. In the analytical model, the $1s$- and $2p$-populations are assumed to undergo unperturbed Rabi cycling induced by the resonant field. The TDSE-calculation reveals, however, that the non-resonant $\omega_1$-field alters the $1s$- and $2p$-population dynamics. Overall, the full calculation validates the signatures of the IATS scheme as described by the analytical model. In addition, the more realistic 2D-TDSE model reveals the influence of high-lying intermediate states inherent to more complex quantum systems and highlights their importance in non-perturbative multiphoton control schemes in general.

\subsection{Counterrotating circularly polarized  pulses}\label{sec:results:CRCP}

\begin{figure}[t]
	\includegraphics[width=\linewidth]{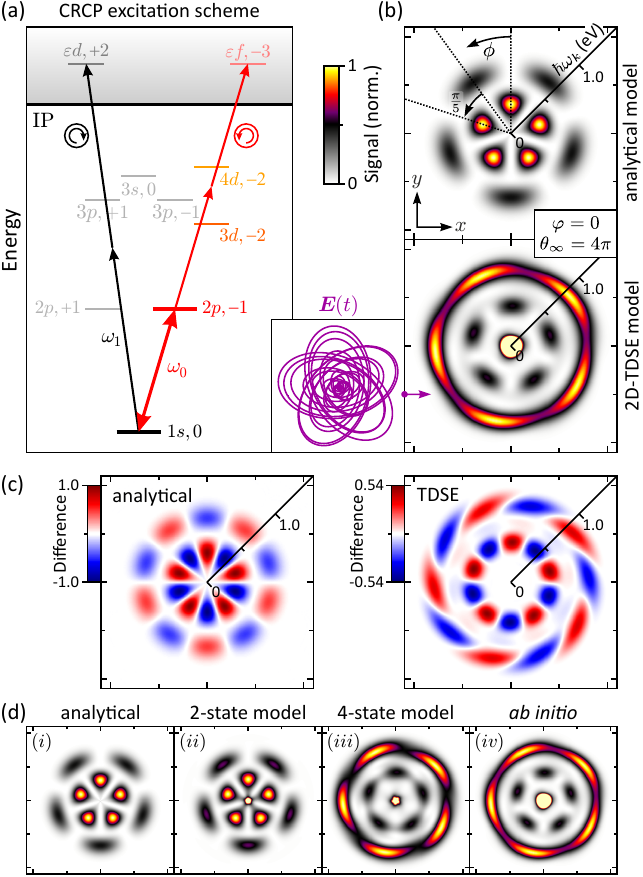}
	\caption{(Color online.) Implementation of the IATS scheme by CRCP \bichrom{2}{3} pulses. The superposition of a $|d,+2\rangle$- and a $|f,-3\rangle$-type photoelectron partial wave packet yields a $c_5$-symmetric 2D PED. The analytical PED is shown in the top frame of (b) and compared to the 2D-TDSE result in the bottom frame. The inset displays the polarization profile of the bichromatic laser electric field $\boldsymbol{E}(t)$. (c) shows the differential PEDs, i.e. the difference between the PEDs obtained for $\varphi=0$ and $\varphi=\pi/2$ for each model. In (d), the analytical (frame $(i)$) and the numerical (frame $(iv)$) PED are compared to the results from the two-state (frame $(ii)$) and the four-state (frame $(iii)$) simulation, respectively.}
	\label{fig:fig3}
\end{figure}

In this section, we demonstrate optical polarization control of the AT doublet, as described in Sec.~\ref{sec:mechanism:control}. For this purpose, we switch the polarization state of the $\omega_1$-field component from RCP to LCP and consider the interaction of the atom with a CRCP \bichrom{2}{3} pulse. All other optical parameters, such as the pulse duration $\Delta t_{\text{fwhm}}$, the field amplitudes $\mathcal{E}_j$ and the relative phase $\varphi=0$, are the same as in the COCP scenario (Sec.~\ref{sec:results:COCP}). The corresponding excitation scheme is depicted in Fig.~\ref{fig:fig3}(a). The analytical model describes the 2D PED in this case as
\begin{align}\label{eq:PED_CRCP}
	\mathcal{P}(\omega_k,\phi)  & = 2\vert\mathcal{A}_2(\delta\omega_k)\vert^2 \bigg[1+\cos\left(5\phi\right)\bigg] \notag \\ 
	& \quad +2\vert\mathcal{A}_2(-\delta\omega_k)\vert^2\bigg[1+\cos\left(5\phi+\pi\right)\bigg],
\end{align}
where the optical phase was set to $\varphi=0$. The physical meaning of the two terms is the same as in Eq.~\eqref{eq:PED_COCP}. However, the angular distribution of the PED, reflecting the angular momentum state of the photoelectron wave packet, is different. As visualized in the top frame of Fig.~\ref{fig:fig3}(b), both AT components exhibit a $c_5$-rotational symmetry. The fast component is rotated against the slow component by an angle of $\pi/5$. Along the $\phi$-direction, the photoelectrons therefore alternate back and forth between the two components. In total, they switch five times in angular intervals of $2\pi/5$, starting with the selective emission of the slow component along $\phi=0$ -- as in the COCP scenario -- and the selective emission of the fast component along $\phi=\pi/5$. \\
The bottom frame of Fig.~\ref{fig:fig3}(b) shows the numerical result from the 2D-TDSE model for comparison. The full calculation agrees qualitatively with the analytical model. In particular, the numerical PED also displays a $c_5$-rotational symmetry. The relative rotation of the two AT components against each other deviates from $\pi/5$ by \SI{0.22}{rad}. This deviation, as well as the enhanced amplitude and the reduced contrast of the fast AT component, are again attributed to the non-resonant intermediate states $3d$ and $4d$, analogous to the discussion of the COCP scenario in Sec.~\ref{sec:results:COCP}. The differential PEDs, obtained by subtracting the calculation results for $\varphi=0$ and $\varphi=\pi/2$, are shown in Fig.~\ref{fig:fig3}(c). The differential PED from the 2D-TDSE model indicates an asymmetry contrast of 54\% achieved in the CRCP scenario. 
To confirm the influence of the intermediate states on the PED, Fig.~\ref{fig:fig3}(d) shows the results from the two-state (panel $(ii)$) and the refined four-state (panel $(iii)$) model simulation, which accurately reproduce the analytical model and the 2D-TDSE model calculation, respectively. Hence, the full calculation validates the analytical approach also in the CRCP scenario, accounting for more realistic conditions. \\
The polarization-control demonstrated here highlights the power of polarization-shaped ultrashort pulses to control besides the energy spectra also the directionality of photoemission processes by careful design of the angular momentum states of the emitted photoelectron wave packets.

\subsection{Linearly polarized pulses}\label{sec:results:LP}

\begin{figure}[t]
	\includegraphics[width=\linewidth]{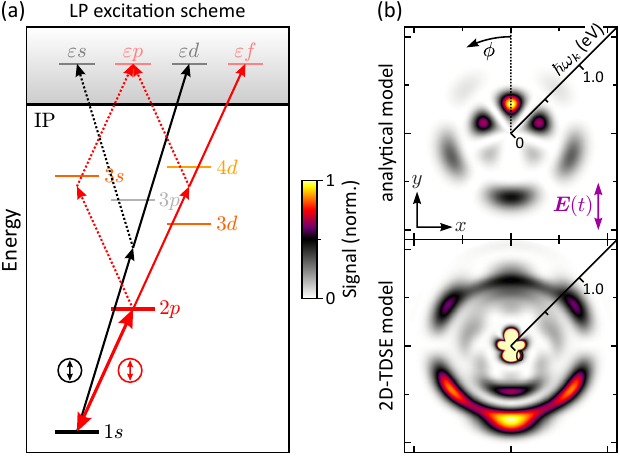}
	\caption{(Color online.) Implementation of the IATS scheme by LP \bichrom{2}{3} pulses. In contrast to the CP scenarios, additional ionization pathways based on $\Delta\ell=-1$ transitions arise in the LP scenario. These pathways are indicated by dotted arrows in (a). (b) If we take only pathways to the $d$- and $f$-type continua into account (solid arrows), as in the CP scenarios, the analytical model (top frame) describes efficient directional control of the AT doublet along the laser polarization direction. However, the full TDSE-calculation (bottom frame), where all ionization pathways are inherently built-in, shows that the selectivity among the AT components in opposite directions is washed out due to the interference with the additional photoelectron partial wave packets from the $s$- and $p$-type continua.}
	\label{fig:fig4}
\end{figure}

The discussion of circularly polarized (CP) scenarios in the previous sections has shown that for a complete description of the 2D PED it is required to consider not only the angle-dependent phases of the angular momentum states but also the energy-dependent ionization phases. To demonstrate the effect of multiple interfering partial wave packets, in this section we discuss the IATS scheme using linearly polarized (LP) bichromatic pulses. Analogously, LP bichromatic microwave pulses with commensurable frequencies were used to study the photoionization of hydrogen und helium Rydberg-atoms \cite{Sirko:2002:PRL:274101,Koch:2003:JPBAMOP:4755}. LP pulses are described by a superposition of two CRCP field components which entails two important implications. First, the angular part of the created photoelectron partial wave packets is described by $Y_m(\phi)=\cos(m\phi)$ in the 2D case. Second, the $\Delta\ell=-1$ transitions are no longer inhibited as in the CP scenarios, which opens up further MPI pathways to additional photoelectron continua. A Grotrian-type excitation scheme including all dipole-allowed MPI pathways for the LP scenario is shown in Fig.~\ref{fig:fig4}(a). The solid arrows indicate the relevant pathways for the CP scenarios based on $\Delta\ell=+1$ transitions. In the LP scenario, these pathways are still favored by the propensity rules \cite{Fano:1985:PRA:617}. For simplicity, we therefore consider only these pathways in our analytical model. Assuming a $y$-polarized \bichrom{2}{3} pulse with $\varphi=0$, we obtain the expression
\begin{align}\label{eq:PED_LP}
	&\mathcal{P}(\omega_k,\phi) = 2\vert\mathcal{A}_2(\delta\omega_k)\vert^2 \bigg[\cos\left(2\phi\right)+\cos\left(3\phi\right)\bigg]^2 \notag \\ 
	 &\quad\quad\;\; +2\vert\mathcal{A}_2(-\delta\omega_k)\vert^2\bigg[\cos\left(2\phi\right)+\cos\left(3\phi+\pi\right)\bigg]^2.
\end{align}
The corresponding 2D PED is visualized in the top frame of Fig.~\ref{fig:fig4}(b). Apart from the angular distribution, it behaves very similarly to that from the COCP scenario in Fig.~\ref{fig:fig2}(a). The slow AT component is observed selectively and with maximum amplitude along $\phi=0$. In the opposite direction, the photoelectrons switch selectively and with maximum efficiency to the fast component. In contrast, the PED from the 2D-TDSE model, shown in the bottom frame of Fig.~\ref{fig:fig4}(b), displays a pronounced asymmetry along the laser polarization direction but no pronounced selectivity among the AT components. Both components are observed with maximum amplitude along $\phi=\pi$. This deviation from the key signature of the IATS scheme is rationalized by the interference of the $d$- and $f$-type photoelectron partial wave packets with additional partial waves from the $s$- and $p$-type continua. The latter are created via the MPI pathways indicated by dotted arrows in Fig.~\ref{fig:fig4}(a). These pathways consist both of $\Delta\ell=+1$ and $\Delta\ell=-1$ transitions. The ionization phases acquired along each pathway are determined by the intermediate resonances, such as the $3s$-state which is accessible in the LP scenario, and the corresponding bound-to-ionic couplings. Hence the different partial wave packets are generally not in phase, canceling the interference mechanism of the IATS scheme. This result highlights that for the complete description of the PED from strong-field MPI the interplay of the angle-dependent phases introduced by the photoelectron angular momentum states and the energy-dependent phases introduced by intermediate states along the MPI pathways need to be taken into account. In view of experimental implementations of the IATS scheme, our simulation results show that the use of CP pulses is advantageous to reduce the number of competing MPI pathways and enable the background-free mapping of the ground- and excited-state dynamics.

\section{Conclusion}\label{sec:conclusion}

In this paper, we proposed a novel scheme for the non-perturbative control of the AT doublet in the photoelectron spectrum from atomic REMPI. The scheme is based on the interference of two AT doublets created by the simultaneous mapping of the ground- and excited-state Rabi dynamics in a strongly driven two-state system using tailored bichromatic femtosecond laser pulses. So far, the AT doublet has been considered an indicator of the bound-state dynamics. In our scheme, the phase of the interfering AT doublets plays the pivotal role and is utilized to control the shape of the resulting AT spectrum. We have shown that, owing to the distinct phase relation between the ground- and excited-state dynamics of the Rabi solution, the interference of the two AT doublets in the energy-dependent photoelectron amplitude is always constructive in the slow component and destructive in the fast component. Unlike the single-color (1+1) REMPI vs. 2PI scheme reported recently in \cite{Nandi:2022:Nature:488,Olofsson:2023:PRR:043017}, the interference condition in the bichromatic IATS scheme is fully controllable by the relative phase between the two colors and the laser polarization state. Depending on the angular momentum state of the photoelectron partial wave packets (determined by the number of photons required for ionization), additional azimuthal phases are generally introduced through the angular part of the wave functions, which renders the interference condition angle-dependent. As a result, the slow AT component is observed selectively in certain directions while in other directions, the photoelectrons switch selectively to the fast component. We provided analytical expressions of the AT doublet for cosine-squared pulses to investigate how the pulse parameters determine the shape of the AT spectrum. However, the scheme is general, i.e., applicable for any kind of pulse with a real-valued envelope irrespective of its shape. \\
Motivated by previous experiments, we demonstrated the IATS scheme on the (1+2) REMPI vs. 2PI of potassium atoms using \bichrom{2}{3} COCP and CRCP pulses. The analytical results were validated against \textit{ab initio} calculations for the interaction of a 2D potassium-like atom with Gaussian-shaped pulses. The full 2D-TDSE calculation confirms the signatures of the IATS scheme and sheds light on the influence of higher lying intermediate states in the multiphoton strong-field control scenarios investigated here. \\
We conclude by considering the implications of the theoretical analysis for the experimental implementation of the proposed scheme. In a general $(1+N)$ REMPI vs. $M$-photon ionization scenario, with $(1+N)\neq M$, the interference condition is angle-dependent. As a consequence, the interference pattern is averaged out in the angle-integrated photoelectron energy-spectrum. (1) Applying differential photoelectron detection techniques, such as velocity map imaging \cite{Eppink:1997:RSI:3477} or COLTRIMS \cite{Doerner:2000:PR:95}, is therefore crucial for the experimental observation of the AT control by the IATS scheme. (2) As shown previously, in this case, the interference mechanism is sensitive to the carrier envelope phase (CEP) \cite{Kerbstadt:2019:NC:658}. The CEP-stabilization \cite{Kerbstadt:2017:OE:12518} or -tagging \cite{Wittmann:2009:NP:357} is hence mandatory, otherwise the interference pattern is averaged out even in the angle-resolved spectrum. (3) Our \textit{ab initio} results suggest that the use of CP instead of LP pulses is advantageous to circumvent competing MPI pathways and ensure the background-free mapping of the ground- and excited-state dynamics. (4) Most importantly, our analytical treatment shows that the interference condition is independent of the field amplitudes. Thus, we expect the proposed AT control scheme to be robust against laser intensity fluctuations as well as focal intensity averaging, which will facilitate the observation of the scheme in the experiment.

\begin{acknowledgments}
	Financial support from the Deutsche Forschungsgemeinschaft (DFG) via the priority program QUTIF (Program No. SPP1840) is gratefully acknowledged.
\end{acknowledgments}

\appendix

\section{Analytical model}\label{app:analytics}

In this section, we derive an analytical expression for the AT doublet in the photoelectron spectrum from atomic $(1+N)$ REMPI vs. non-resonant $M$-photon ionization. Using a bichromatic field with suitably chosen central frequencies $\omega_0$ and $\omega_1$, the two ionization processes map the Rabi dynamics of two strongly coupled bound-states, labeled $1s$ and $2p$, into the same energy window of the photoelectron continuum. The $\omega_0$-field component is considered to be resonant with the atomic transition $1s\rightarrow2p$ by setting the eigenenenergies of the bound-states to $\hbar\omega_{1s}=0$ and $\hbar\omega_{2p}=\hbar\omega_0$. For both colors, we assume a cosine-squared envelope $f(t)$ of the electric field $E_n^+(t)=\mathcal{E}_nf(t)e^{i\omega_n t}$ ($n=0,1$) similar to \cite{Wu:2013:PRA:043416}:
\begin{equation}\label{eq:envelope}
f(t) = 
\begin{cases}
\cos^2 \left( \frac{\pi t}{\Delta t} \right) & ; ~\text{if} ~ -\frac{\Delta t}{2} \leq t \leq \frac{\Delta t}{2}   \\
0 &  ; ~\text{else}
\end{cases}.
\end{equation}
$\Delta t$ is the footprint pulse duration. The Rabi solution for the population amplitudes of the $1s$ ground state and the $2p$ excited state of the resonantly driven two-state system reads \cite{Rabi:1937:PR:652,Shore:1990a}
\begin{equation}\label{eq:Rabi_app}
c_{1s}(t) = \cos \left[ \frac{\theta(t)}{2} \right] \quad \text{and} \quad c_{2p}(t) = i \sin \left[ \frac{\theta(t)}{2} \right],
\end{equation}  
with the time-dependent pulse area $\theta(t)$ for the cosine-squared pulse
\begin{align}
\theta(t) & = \frac{\mu_0\mathcal{E}_0}{\hbar} \intop_{- \frac{\Delta t}{2}}^{ t} f(t') \, dt' \notag\\
& = \frac{\theta_\infty}{2\pi} \left[\sin\left( \frac{2\pi t}{\Delta t} \right) + \frac{2\pi t}{\Delta t} + \pi \right]. \label{eq:pulsearea}
\end{align}  
The last expression is valid for $-\frac{\Delta t}{2}\leq t \leq \frac{\Delta t}{2}$, and $\theta_\infty=\theta\!\left(\frac{\Delta t}{2}\right)=\frac{\mu_0\mathcal{E}_0}{2\hbar}\Delta t$ denotes the final pulse area.

\subsection{Photoionization from the $2p$ excited state}

\begin{figure}[t]
	\includegraphics[width=\linewidth]{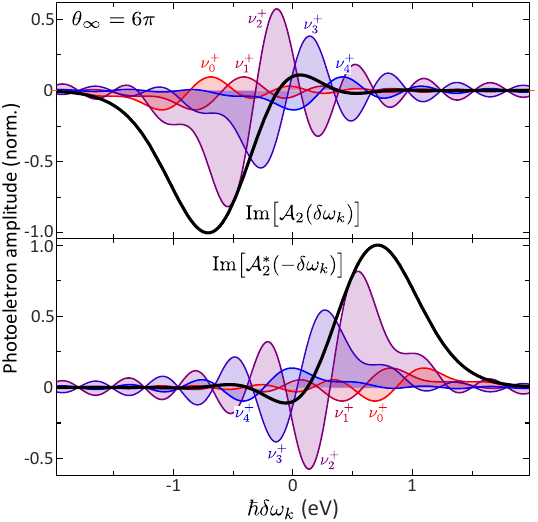}
	\caption{(Color online.) Illustration of the photoelectron amplitude $a_{2p}(\delta\omega_k)$ according to Eq.~\eqref{eq:pe_spec_exc}, from (1+2) REMPI ($N=2$) by a cosine-squared pulse with pulse area $\theta_\infty=6\pi$. The top frame shows the slow AT component (bold black line) described by the partial amplitude $\mathcal{A}_2(\delta\omega_k)$, which is purely imaginary in this case. The colored lines (shaded backgrounds) display the corresponding decomposition into Anger functions $\boldsymbol{\mathrm{J}}_{\nu_j^+(\delta\omega_k)}$ ($j=0,..,4$). The bottom frame shows the fast AT component determined by $\mathcal{A}_2^\ast(-\delta\omega_k)$. }
	\label{fig:app1}
\end{figure}

Since the bound-to-ionic couplings are generally much weaker than the couplings in the bound-state system, the photoionization is described by time-dependent perturbation theory \cite{Meier:1994:PRL:3207,Meshulach:1999:PRA:1287,Wollenhaupt:2003:PRA:015401}. We consider non-resonant $N$-photon ionization from the $2p$-state to a continuum state with energy $\hbar(\omega_k+\omega_\text{IP})$, where $\hbar\omega_k$ is the photoelectron kinetic energy and $\hbar \omega_\text{IP}$ is the IP. Then the energy-dependent photoelectron amplitude is given by \cite{Meier:1994:PRL:3207,Meshulach:1999:PRA:1287}
\begin{equation}\label{eq:peformula}
a_{2p}(\delta\omega_k) \propto \intop_{- \frac{\Delta t}{2}}^{\frac{\Delta t}{2}} c_{2p}(t) \,  f^N(t) \, e^{i \delta\omega_k t} \,  dt,
\end{equation}
with $\delta\omega_k=\omega_k+\omega_\text{IP}-(N+1)\omega_{2p}$ being the $N$-photon detuning of the $\omega_0$-field component ($\omega_0=\omega_{2p}$) relative to the transition from the $2p$-state to the continuum state. Using Euler's formula along with the binomial theorem, the $N$-th order pulse envelope in Eq.~\eqref{eq:peformula} is written as	
\begin{align}
f^N(t) = \cos^{2N} \left(\frac{\pi t}{\Delta t} \right) = \frac{1}{2^{2N}} \sum_{j=0}^{2N} \begin{pmatrix}
2N \\ 
j
\end{pmatrix} 
e^{\pm i \frac{\pi t}{\Delta t} [2(N-j)]}.
\end{align}
Because the term $(N-j)$ runs symmetrically from $-N$ to $+N$, either the plus or the minus sign of the exponent can be selected, which we will make use of in the next step. Expanding the excited-state amplitude $c_{2p}(t)$ in Eq.~\eqref{eq:Rabi_app} into exponential functions and rearranging the terms, the integrand of Eq.~\eqref{eq:peformula} becomes
\begin{align}\label{eq:integrand}
& c_{2p}(t) \,  f^N(t) \, e^{i \delta\omega_k t} 
=  i \, \sin \left[ \frac{\theta(t)}{2} \right]  \cos^{2N} \left( \frac{\pi t}{\Delta t} \right) \, e^{i \delta\omega_k t} \notag \\
& \hspace{2.75cm}= \frac{1}{2^{2N+1}} \sum_{j=0}^{2N} \begin{pmatrix}
2N \\ 
j
\end{pmatrix} \notag \\
& \quad\quad\quad\quad\quad \times\left[ e^{i  \frac{\theta_\infty}{4}  } \, e^{i (\delta\omega_k +  \Delta_j )t } \, e^{i \frac{\theta_\infty}{4\pi} \sin\left( \frac{2\pi t}{\Delta t} \right) }   \right.   \notag \\
& \quad\quad\quad\quad\quad\;\;\, \left.-  e^{-i  \frac{\theta_\infty}{4} } \, e^{i(\delta\omega_k - \Delta_j)t } \, e^{-i \frac{\theta_\infty}{4\pi} \sin\left( \frac{2\pi t}{\Delta t} \right) } \right],
\end{align} 
where we introduced the shorthand notation $\Delta_j =\frac{2\pi}{\Delta t} \left(N - j + \frac{\theta_\infty}{4\pi}\right)$. Substituting $\xi = \frac{2\pi t}{\Delta t}$, the photoelectron amplitude in Eq.~\eqref{eq:peformula} is rewritten as
\begin{align}\label{eq:app_exc_ampl_long}
a_{2p}(\delta\omega_k) \propto & \frac{\Delta t}{4^{N+1}} \, \sum_{j=0}^{2N} \begin{pmatrix}
2N \\ 
j
\end{pmatrix} \notag \\
& \times   \intop_{- \pi}^{ \pi}  \left[  e^{i  \frac{\theta_\infty}{4} } \,  e^{-i \nu_j^{+}(\delta\omega_k) \xi} \, e^{i \frac{\theta_\infty}{4\pi} \sin( \xi) } \right. \notag  \\    
& \quad\quad \left. -   e^{-i \frac{\theta_\infty}{4} } \, e^{-i \nu_j^{-}(\delta\omega_k) \xi} \, e^{-i \frac{\theta_\infty}{4\pi} \sin(\xi) }  \right] d\xi ,
\end{align}
where 
\begin{align}
	\nu_j^\pm (\delta\omega_k) & = -\frac{\delta\omega_k \pm \Delta_j}{2\pi}  \Delta t \label{eq:app:nu_symmetry_rel}\\
	& = -\left[\pm\left(N-j+\frac{\theta_\infty}{4\pi}\right)+\frac{\delta\omega_k\Delta t}{2\pi}\right].
\end{align}
By using the real-valued Anger-function $\boldsymbol{\mathrm{J}}_\nu$ of the order $\nu$ defined as \cite{Gradshteyn:2014}
\begin{equation}
\boldsymbol{\mathrm{J}}_{\nu}(\beta) = \frac{1}{2 \pi} \, \intop_{- \pi}^{ \pi} e^{\pm i \nu \xi} e^{\mp i \beta \sin(\xi)} d\xi,
\end{equation}
with $\mathrm{Re}[\beta]>0$, and exploiting the relation
\begin{equation}
	\nu_j^+ (-\delta\omega_k)=-\nu_j^- (\delta\omega_k)
\end{equation}
following from Eq.~\eqref{eq:app:nu_symmetry_rel}, we find
\begin{align}\label{eq:app:S_e_long}
a_{2p}(\delta\omega_k) \propto & \;\frac{\Delta t}{4^{N+1}} \, \sum_{j=0}^{2N} \begin{pmatrix}
2N \\ 
j
\end{pmatrix} \, \notag \\ 
& \quad\quad\times \left[  e^{i \frac{\theta_\infty}{4} } \boldsymbol{\mathrm{J}}_{\nu_j^+ (\delta\omega_k)}\left(\frac{\theta_\infty}{4\pi} \right) \right. \notag \\
& \quad\quad\;\;\, \left. - e^{-i \frac{\theta_\infty}{4} }  \boldsymbol{\mathrm{J}}_{\nu_j^+ (-\delta\omega_k)}\left(\frac{\theta_\infty}{4\pi} \right)  \right] .
\end{align}
Introducing the photoelectron partial amplitude by
\begin{align}
\mathcal{A}_{N}(\delta\omega_k) &= \frac{\Delta t}{4^{N+1}} \, e^{i \frac{\theta_\infty}{4}  } \, \sum_{j=0}^{2N} \begin{pmatrix}
2N \\ 
j
\end{pmatrix} \,   \boldsymbol{\mathrm{J}}_{\nu_j^+ (\delta\omega_k)}\left(\frac{\theta_\infty}{4\pi} \right),
\end{align}
the total photoelectron amplitude can be written in a compact manner as
\begin{align}\label{eq:pe_spec_exc}
a_{2p}(\delta\omega_k) &\propto   \mathcal{A}_{N}(\delta\omega_k) - \mathcal{A}_{N}^*(-\delta\omega_k).
\end{align}
Exemplarily, the partial amplitudes $\mathcal{A}_2(\delta\omega_k)$ and $\mathcal{A}_2^*(-\delta\omega_k)$, constituting the photoelectron amplitude from $(1+2)$ REMPI via the $2p$-state by a $6\pi$-pulse, are visualized in Fig.~\ref{fig:app1} (bold black lines) together with their decomposition into the Anger functions $\boldsymbol{\mathrm{J}}_{\nu_j^+(\delta\omega_k)}$ (colored lines with shaded backgrounds). We see that the first term in Eq.~\eqref{eq:pe_spec_exc} describes the slow component of the AT doublet while the second term describes the fast component. Note that Eq.~\eqref{eq:pe_spec_exc} is valid for any real-valued pulse shape $f(t)$. By defining
\begin{equation}
c(t) = f^N(t) \, e^{i\frac{\theta(t)}{2}}
\end{equation}
and considering its Fourier transform $\mathcal{F}$ as a function of the variable $\delta\omega_k$ 
\begin{equation}
\mathcal{A}_N(\delta\omega_k) = \mathcal{F}\big[ c(t) \big](\delta\omega_k),
\end{equation}
implying 
\begin{equation}
\mathcal{A}^*_N(-\delta\omega_k) = \mathcal{F}\big[ c^*(t) \big](\delta\omega_k),
\end{equation}
we see that the photoelectron amplitude resulting from $N$-photon ionization of the excited state in a Rabi oscillating system always has the form given in Eq.~\eqref{eq:pe_spec_exc}. Equation~\eqref{eq:pe_spec_exc} describes the photoelectron amplitude as a superposition of the two partial amplitudes with a respective phase of $\pm\theta_\infty/4$. At the center of the AT doublet, at $\delta\omega_k = 0$, both partial amplitudes have the same modulus. Therefore we find
\begin{align}\label{eq:app:Se_varpi0}
a_{2p}(\delta\omega_k = 0) &\propto i \sin\left(\frac{\theta_\infty}{4}\right),
\end{align}
i.e., the amplitude varies periodically with the pulse area $\theta_\infty$ determined by the field amplitude $\mathcal{E}_0$. In addition, Eq.~\eqref{eq:app:Se_varpi0} shows that the amplitude at the center of the AT doublet is always imaginary (or zero). \\
Eventually, we consider the full photoelectron wave function created by $(1+N)$ REMPI via the $2p$-state using a CP pulse. In the 2D case studied here, the angular part of the wave function is given by the single-valued eigenfunction of the angular momentum, i.e., the circular harmonic $Y_m(\phi) \propto e^{i m \phi}$ \cite{Parfitt:2002:JMP:4681a}. For $(1+N)$-photon ionization by an RCP pulse, the angular momentum quantum number reads $m=-(1+N)$, hence we obtain
\begin{align}\label{eq:wf_spec_exc}
\psi_{2p}(\omega_k,\phi) &= a_{2p}(\delta\omega_k) Y_m(\phi) \notag \\ & \propto   \left[ \mathcal{A}_N(\delta\omega_k) - \mathcal{A}_N^*(-\delta\omega_k) \right] e^{-i (1+N) \phi}.
\end{align}

\subsection{Photoionization from the $1s$ ground state}

The derivation of the photoelectron amplitude created by non-resonant $M$-photon ionization from the $1s$ ground state by the $\omega_1$-field component is fully analogous to the procedure described above. By inserting the ground state amplitude $c_{1s}(t)$ from Eq.~\eqref{eq:Rabi_app} into the photoelectron integral in Eq.~\eqref{eq:peformula} and allowing for an additional optical phase $\varphi$, we find
\begin{equation}\label{eq:amp_spec_grou}
a_{1s}(\delta\omega_k) = \left[ \mathcal{A}_M(\delta\omega_k) + \mathcal{A}_M^*(-\delta\omega_k) \right] e^{-i M \varphi}.
\end{equation}
The essential difference between Eq.~\eqref{eq:amp_spec_grou} and Eq.~\eqref{eq:pe_spec_exc} is the sign of the fast AT component, which results from the cosinusoidal and sinusoidal behavior of the $1s$- and $2p$-amplitude, respectively (see Eq.~\eqref{eq:Rabi_app}). This difference is the key feature of the IATS scheme. The 2D photoelectron wave function from $M$-photon ionization of the $1s$-state by a CP pulse thus takes the form
\begin{align}\label{eq:pe_spec_grou}
\psi_{1s}(\omega_k,\phi) &\propto  \left[   \mathcal{A}_M(\delta\omega_k) +  \mathcal{A}_M^*(-\delta\omega_k) \right] e^{-i M (\varphi\mp\phi)},
\end{align}
where the minus (plus) sign corresponds to LCP (RCP) ionization. Note that, in this case, the photoelectron amplitude at the center of the AT doublet is always real-valued: $a_{1s}(\delta\omega_k = 0) \propto \cos(\theta_\infty/4)$.

\subsection{Interference of Autler-Townes doublets}

By designing the central frequencies of the bichromatic field such that $(1+N)\omega_0 = M\omega_1$, the AT doublets from ground- and excited state-ionization are mapped into the same energy window of the continuum and interfere. Motivated by the discussion in the main text (see Sec.~\ref{sec:mechanism:scheme}), we describe the interference for the case $N=M$, where the same number of photons are required for the ionization from both states. However, it is straightforward to extend the formalism to a general $(1+N)$ REMPI vs. $M$-photon ionization scenario. Assuming equal amplitudes of the two partial wave packets, by suitable choice of the field amplitudes $\mathcal{E}_n$, the coherent superposition wave function $\psi_{tot}(\omega_k,\phi)$ reads
\begin{align}\label{eq:app_superpos}
\psi_{tot}(\omega_k,\phi)&= \psi_{e,1s}(\omega_k,\phi) + \psi_{e,2p}(\omega_k,\phi) \notag \\ & \propto  \mathcal{A}_N(\delta\omega_k) \left( 1 +  e^{i (\sigma \phi - N\varphi)}  \right) \notag \\ & \quad -  \mathcal{A}_N^*(-\delta\omega_k) \left( 1 + e^{i (\sigma \phi - N\varphi + \pi)} \right).
\end{align}
The parameter ${\sigma = 1+N\pm N}$ accounts for the polarization state of the bichromatic field. The plus (minus) sign corresponds to a CRCP (COCP) pulse. The corresponding 2D photoelectron momentum distribution $\mathcal{P}(\omega_k,\phi) = \vert \psi_{tot}(\omega_k,\phi) \vert^2$ reads
\begin{align}
\label{eq:2D_PED}
\mathcal{P}(\omega_k,\phi)  &\propto 2 \vert \mathcal{A}_N(\delta\omega_k) \vert^2 \bigg[ 1 + \cos \left( \sigma \phi - N\varphi \right) \bigg] \notag \\ 
& \quad+ 2 \vert \mathcal{A}_N(-\delta\omega_k) \vert^2 \bigg[ 1 + \cos \left( \sigma \phi + \pi - N\varphi \right) \bigg] \notag \\ 
& \quad + 4 \vert \mathcal{A}_N(\delta\omega_k) \mathcal{A}_N(-\delta\omega_k) \vert \notag \\
& \quad\quad\quad\quad \times \sin\left(\theta_\infty/2\right) \sin\big( \sigma \phi - N\varphi \big).
\end{align}
The last term in Eq.~\eqref{eq:2D_PED}, i.e., the mixing contribution can be neglected for a sufficiently small overlap of the two AT components. In addition, the mixing term vanishes exactly if the pulse area  $\theta_\infty$ equals even multiples of $2\pi$. The first and second term of $\mathcal{P}(\omega_k,\phi)$ describe two $\sigma$-fold rotationally symmetric contributions which are rotated against each other about an angle of $\pi/\sigma$. The overall rotation of the PED is controllable by the relative optical phase $\varphi$ between the two components of the bichromatic pulse. Finally, we note that while Eq.~\eqref{eq:2D_PED} was derived assuming a cosine-squared pulse, the scheme is more generally applicable for any pulse with a real-valued envelope irrespective of its shape.

\section{2D-TDSE model}\label{app:ab_initio}

The numerical methods used in our 2D-TDSE model have been described in detail elsewhere \cite{Bayer:2020:PRA:013104,Bayer:2023:PRA:033111}. Briefly, we solve the 2D-TDSE in the dipole approximation and length gauge 
\begin{equation}\label{eq:tdse}
	i\hbar\frac{\partial}{\partial t}\psi(\boldsymbol{r},t)=\left[-\frac{\hbar^2}{2m_e}\Delta+V(r)+e\,\boldsymbol{r}\cdot\boldsymbol{E}(t)\right]\psi(\boldsymbol{r},t)
\end{equation}
for a single active electron with mass $m_e$ and charge $-e$ in the soft-core Coulombic potential \cite{Sprik:1988:JCP:1592,Shin:1995:JCP:9285,Erdmann:2004:JCP:9666a} 
\begin{equation}\label{eq:potential}
	V(r)=-\frac{z e^2}{4\pi\varepsilon_0}\frac{\mathrm{erf}(r/a)}{r}.
\end{equation}
The electron interacts with a polarization-shaped laser electric field $\boldsymbol{E}(t)$ which is described by the real part of its positive frequency analytic signal $\boldsymbol{E}(t)=\mathrm{Re}[\boldsymbol{E}^+(t)]$ and represented in the spherical basis as \\
\begin{equation}\label{eq:field_CP}
	\boldsymbol{E}^+(t) = f(t) \big(\mathcal{E}_0 e^{i\omega_0 t}\,\boldsymbol{e}_{q_1}+\mathcal{E}_1 e^{i(\omega_1 t+\varphi)}\,\boldsymbol{e}_{q_2}\big),
\end{equation}
with $q_n=\pm1$ for LCP and RCP light, respectively, and $\boldsymbol{e}_{\pm1}=(\boldsymbol{e}_x\mp i\boldsymbol{e}_y)/\sqrt{2}$. For LP light polarized in $y$-direction, the spherical unit vectors  $\boldsymbol{e}_{q_n}$ are both replaced by the cartesian unit vector $\boldsymbol{e}_y=i(\boldsymbol{e}_{+1}-\boldsymbol{e}_{-1})/\sqrt{2}$. \\
The atom is initially prepared in the ground state. The ground state wave function is refined by imaginary-time propagation \cite{Tal-Ezer:1986:CPL:223}. Subsequently, the wave function $\psi(\boldsymbol{r},t)$ is propagated on a discrete spatial grid using a Fourier-based split operator technique \cite{Feit:1982:JCP:412a}. Non-physical reflections at the spatial boundaries are minimized using absorbing boundary conditions \cite{Kosloff:1986:JCP:363,Santra:2006:PRA:034701}. After the laser-atom interaction, the wave function is propagated until the free part $\psi_f(\boldsymbol{r},t)$, i.e. the photoelectron wave packet, has detached from the bound part but not yet reached the absorbing boundaries. At this time, $t=t_f$, the photoelectron wave function is separated from the bound part by application of a circular splitting filter \cite{Heather:1987:JCP:5009}. Fourier transformation of the free part yields the 2D photoelectron momentum distribution $\mathcal{P}(\boldsymbol{k})=\big|\mathcal{F}[\psi_f(\boldsymbol{r},t_f)](\boldsymbol{k})\big|^2$. Calibration of $\mathcal{P}(\boldsymbol{k})$ according to the relation $\boldsymbol{k}\rightarrow\omega_k=\frac{\hbar}{2m_e}\boldsymbol{k}^2$ finally yields the 2D PED $\mathcal{P}(\omega_k,\phi)$.


%

\end{document}